\documentclass{aastex}
\usepackage[twocolumn]{emulateapj5}

\def\ltsima{$\; \buildrel < \over \sim \;$}
\def\simlt{\lower.5ex\hbox{\ltsima}} 
\def\gtsima{$\; \buildrel > \over \sim \;$}
\def\simgt{\lower.5ex\hbox{\gtsima}} 

\usepackage{epsfig}
\usepackage{rotating}
\usepackage{apjfonts}

\submitted{ApJ, accepted}

\shorttitle{Optical Hotspot Detections in 3CR Quasars} 
\shortauthors{Cheung et al.}

\begin{document}

\title{Discovery of Optical Emission in the Hotspots of Three 3CR 
Quasars: High-Energy Particle Acceleration in Powerful Radio Hotspots}

\author{C.~C. Cheung\altaffilmark{1,2}, J.~F.~C. Wardle, T. Chen}

\affil{Department of Physics, MS~057, Brandeis University,
Waltham, MA 02454}

\altaffiltext{1}{Jansky Postdoctoral Fellow; National Radio Astronomy
Observatory}

\altaffiltext{2}{Address: MIT, Kavli Institute for Astrophysics \& Space
Research, 77 Massachusetts Ave., Cambridge, MA 02139;
ccheung@space.mit.edu}

\begin{abstract}

Archival Hubble Space Telescope WFPC2 images were used to search for
optical emission associated with the radio jets of a number of luminous
quasars.  From this search, we report new optical hotspot detections in
the well-known blazar 3C~454.3 and the lobe-dominated quasars 3C~275.1 and
3C~336. We also find optical emission in the vicinity of the hotspot in
3C~208, but believe this is a chance alignment.  Optical emission from the
arcsecond-scale jet in 3C~454.3 is also detected. Multi-frequency archival
radio data from the VLA and MERLIN are analyzed, and the synchrotron
spectra of these high-power hotspots are presented.  We estimate that
their break frequencies are in the range of $10^{10}-10^{11}$ Hz, with
large uncertainties due to the wide gap in frequency coverage between the
radio and optical bands.  We also calculate their equipartition magnetic
fields, and find that the anti-correlation between break frequency and
magnetic field found by Brunetti et al. for lower power hotspots extends
to these high power hotspots. This supports their model of hot-spots based
on shock acceleration and synchrotron losses. 

\end{abstract}


\section{Introduction}

The production of very high energy particles via diffusive shock
acceleration has applications to many astrophysical problems (e.g.,
Blandford and Eichler 1987). For instance, it is believed to be
responsible for the production of TeV energy particles observed in
supernova remnants, and in the very inner regions of active galactic
nuclei (Catanese \& Weekes 1999). 

Another important site of high energy particle acceleration is in the
terminal hotspots of powerful extended extragalactic radio sources. Their
synchrotron spectra are observed from radio to optical frequencies (e.g.,
Meisenheimer et al. 1989), with the most energetic radiation requiring
electrons with Lorentz factors, $\gamma$$\simgt 10^{5}$, which must be
accelerated in-situ. The recent rejuvenation of interest in the physics of
hotspots is due mainly to the launch of the Chandra X-ray Observatory. 
Chandra's sub-arcsecond resolution imaging capability and highly sensitive
detectors allow for a unique probe of their physical conditions -- e.g.,
magnetic field strength, when the X-ray emission can be attributed to
synchrotron self-Compton radiation, and energetics (e.g., Hardcastle et
al. 2002). 

Prior to the launch of Chandra, most studies of hotspots concentrated on
high-resolution radio to optical mapping of their spectra, which can in
principle, constrain their key physical parameters -- synchrotron break
frequency, magnetic field, electron mean free path. However, the higher
frequency observations were limited to ground-based optical and infrared
instruments, and met with mixed success (e.g., Meisenheimer et al. 1989,
1997). 

Recently, Brunetti et al. (2003) used the VLT and successfully detected
optical counterparts to 6 out of 10 radio hotspot targets (one
``non-detected'' hotspot was limited mainly by confusion with a field
source).  Their targets were lower radio power hotspots than the more
powerful ones which were previously searched for optical emission (e.g.,
Meisenheimer et al. 1989, 1997). The lower power hotspots suffer less
losses, so are sites of more efficient particle acceleration to the high
energies ($\gamma$$\sim$10$^{5}$) which are required to produce optical
emission (Prieto et al. 2002).  Conversely, the higher radio power
hotspots, i.e. with larger magnetic fields, suffer greater (synchrotron
and inverse Compton) losses, so show cutoffs in their spectra at lower
frequencies -- see the comparison between the spectra of the archetype
high and low power hotspots in Cygnus A and 3C~445, respectively, in
Figure~2 of Brunetti et al. (2003). 

In this scenario, there is a strong magnetic field dependence of the break
frequency in their synchrotron spectra ($\nu_{\rm b}\propto$$B^{-3}$) and
observations of the lower power hotspots support this picture (Brunetti et
al. 2003). It is important to test this proposal, by extending these
results to higher power hotspots (with greater $B$-fields). However,
because their spectra are expected to cut off at lower energies, the
higher power hotspots are faint in the optical, and can only be detected
in deep Hubble Space Telescope (HST) images. 

The HST has demonstrated an unique capability of being able to detect
faint optical counterparts to radio jet/hotspot features. This has proved
to be important for the interpretation of the bright X-ray jets detected
by Chandra (Tavecchio et al. 2000; Celotti, Ghisellini, \& Chiaberge 2001;
Sambruna et al. 2004). We were motivated by these work and after a
successful detection of a bright optical knot in the well-studied quasar
3C~279 from archival HST data (Cheung 2002; also motivated by the Chandra
results on this object; Marshall et al. 2003), we initiated a similar
search for extended optical features in other radio sources. Our search
yielded optical detections of hotspots in three double-lobed 3CR quasars
(3C~263, 3C~275.1, and 3C~336), the hotspot and jet knots in the blazar
3C~454.3, one tentative hotspot detection in 3C~208, and confirmation of
the previously detected optical jet in PKS~1229--021 (Le Brun et al.
1997).  The optical hotspot detection in 3C~263 has since been
independently published by Hardcastle et al. (2002) who analyzed it in
detail along with the Chandra detection. Neither 3C~263 or PKS~1229--021
will be discussed further. In this paper, we focus our discussion on the
results on the other hotspots in light of the work of Brunetti et al.
(2003). Initial results on 3C~454.3 and PKS~1229--021 were presented in
Cheung et al. (2003). 

In order to facilitate a comparison of our new optical detections with
those of Brunetti et al. (2003), we assume $H_{0}$ = 50 km s$^{-1}$
Mpc$^{-1}$ and $q_{0}$ = 0.5 throughout this paper.

\section{Search Strategy, Data and Analysis}

The on-line Multimission Archive at the Space Telescope
(MAST\footnote{\label{footnote:mast}http://archive.stsci.edu/}) Science
Institute was utilized to identify candidates for this study. MAST allowed
us to identify archival HST images of quasars by their radio flux by
referencing the catalog of Veron-Cetty \& Veron (1996).  For this study,
we chose an integrated 5 GHz flux greater than 0.5 Jy in order to ensure
that the source contained radio features bright enough for convincing
optical detections (assuming a typical range of radio-to-optical spectral
indices of 0.8--1, where $F_{\nu}\propto\nu^{-\alpha}$;  e.g., Sambruna et
al. 2004). Our query in Summer 2002 resulted in about 300 sources that had
data obtained with the various imaging CCDs aboard HST. We focused on
about 1/3 of these sources which had deep WFPC2 exposures (on both the
high resolution planetary camera (PC) chip and any of the three lower
resolution wide field (WF) chips) using one or more orbits of total
exposure time. It was in this type of exposure that the optical jet in
PKS~0637--752 was detected (Chartas et al. 2000; Schwartz et al. 2000).
Lastly, we cross-referenced these candidates with known arcsecond-scale
radio jets, aided by the catalog of Liu \& Zhang (2002) which led to about
a dozen sources searched. This simple procedure was adopted in the
interest of efficiently finding as many previously unidentified optical
jets and hotspots as possible, with little intention of studying any
complete sample. 

\begin{sidewaystable*}
\footnotesize
\centerline{\bf Summary of Multi-Telescope Archival Data}
\begin{center}
\begin{tabular}{lcccccc}
\hline \hline
Source                &Instrument            
&Date                 &Frequency            
&Exp Time             &Program
&Observer or Reference\\
(1)                   &(2)
&(3)                  &(4)
&(5)                  &(6)
&(7)\\
\hline
3C~208
&MERLIN                   & 14 Aug 1993&1.413+1.658           & 14 hr& --    & --\\
&VLA A-configuration      & 13 Dec 1992&8.44                  & 3500 & AL280 & S. Garrington\\
&{\it HST} WFPC2 WF3/F675W& 23 Mar 1997&4.463$\times$10$^{5}$ & 9400 & 6491  & A. Stockton\\
\\
3C~275.1
&MERLIN                   & 07 Aug 1993&1.413+1.658           & 12 hr& --    & --\\
&VLA A-configuration      & 06 Mar 1982&4.89                  & 2340 & WARD  & J.~F.~C. Wardle\\
&VLA A-configuration      & 13 Dec 1992&8.44                  & 3490 & AL280 & S. Garrington\\
&{\it HST} WFPC2 PC1/F675W& 25 Jul 1995&4.463$\times$10$^{5}$ & 1800 & 5978 & S. Rawlings\\
\\
3C~336
&MERLIN                   & 13 Aug 1993&1.413+1.658           & 15 hr& --    & --\\
&VLA A-configuration      & 06 Mar 1982&4.89                  & 2310 & WARD  & J.~F.~C. Wardle\\ 
&VLA A-configuration      & 13 Dec 1992&8.44                  & 4440 & AL280 & S. Garrington\\
&{\it HST} WFPC2 WF2/F814W& 20 Jul 2000&3.755$\times$10$^{5}$ & 6600 & 8220  & M. Franx\\  
&{\it HST} WFPC2 WF2/F702W& 12-14 Sep 1994&4.334$\times$10$^{5}$ & 24000 &5304 &\citet{ste97}\\
&{\it HST} WFPC2 WF3/F622W& 20 Dec 1994&4.846$\times$10$^{5}$ & 7200 & 5401  & \citet{rid97}\\
&{\it HST} WFPC2 WF2/F606W& 20 Jul 2000&4.993$\times$10$^{5}$ & 6600 & 8220  & M. Franx\\
\\
3C~454.3
&VLA A-configuration      & 31 Jan 1985&1.50                  & 10860& AC120 & R. Perley \\
&VLA A-configuration      & 31 Jan 1985&4.86                  & 10490& AC120 & R. Perley \\
&VLA B-configuration      & 06 Jun 2002&8.46                  & 860  & AC641B& C.~C. Cheung \\
&VLA BnA-configuration    & 28 May 2002&22.46                 & 3850 & AC641A& C.~C. Cheung \\
&{\it HST} WFPC2 WF2/F702W& 06 Dec 1997&4.333$\times$10$^{5}$ & 2100 & 6619  & \citet{che01} \\
\hline \hline
\end{tabular}
\caption[Sample Summary]{
\label{table-1}
Column (1) specifies the common name of the sources.\\
Column (2) indicates the telescope and instrument/configuration used.
The HST WFPC2 CCD chip and filters used are indicated. \\
Column (3) is the UT date of the archival observation. \\
Column (4) is the observing frequency in GHz. \\
Column (5) is the time on source in seconds unless otherwise indicated. 
The integration time for the radio 
data correspond to the unedited data.\\
Column (6) are the VLA or HST program codes/numbers.\\
Column (7) is reference for observation or observer name if we could not locate the reference.
}
\end{center}
\end{sidewaystable*}

\subsection{Archival Hubble Space Telescope WFPC2 Data}

Final combined and cosmic-ray removed HST WFPC2 images were available from
the WFPC2 Associations website (see footnote~\ref{footnote:mast}).  Where
there was no WFPC2 Associations image available, we downloaded the
original exposures from the HST archive and combined them with the CRREJ
task in IRAF\footnote{IRAF is distributed by the National Optical
Astronomy Observatories, which are operated by the Association of
Universities for Research in Astronomy, Inc., under cooperative agreement
with the National Science Foundation.}. The final data used for this study
are listed in Table~\ref{table-1}. 

Our identifications were made by overlaying high quality radio maps (see
below) in digital form with the optical data.  HST has the well-known
advantage over ground based telescopes of being very sensitive to faint
point sources, so this procedure was essential to distinguish between
field sources and radio jet related optical features.  We also processed
the optical images with an unsharped masking filter where smoothed
versions of the original images were subtracted from the originals. This
additional processing helped us to identify the nearest lying features to
the bright optical nuclei (within $\sim$1\arcsec) in previous work with
similar data (Cheung 2002; Sambruna et al. 2004), and led here to the
successful detection of optical emission in the jet of 3C~454.3
(\S\ref{sec:3c454.3}). 

Because of this heightened sensitivity to point sources, many of the
imaging fields were littered with sources unassociated with the radio
source which could be mistakened for optical counterparts to jet knots. 
Therefore, in order to get a qualitative sense of the probability of a
false detection, we made two tests.  First, we repositioned the radio core
onto several of the other prominent optical sources (field stars,
galaxies) in the images and did not find any objects aligning with peaks
in the radio jets within 1-2 WFPC2 pixels (i.e., to better than
$\sim$0.1\arcsec\ to 0.2\arcsec ; each of our reported detections satisfy
this).  Second, we simply rotated the radio image in a random fashion with
respect to the optical frame and again, did not find any convincing
alignments with any knots in the radio jet.  We therefore believe that the
detections claimed here are real and are not chance alignments of field
sources. 

Counts were measured from the optical hotspots, which were well-isolated
from any other bright field sources, using circular apertures with the
QPHOT task in IRAF. This gives essentially the counts in an infinite
radius aperture by estimating the background signal in circular annuli
around the source.  Counts in the features close to the bright nucleus in
3C~454.3 were measured with square apertures. We estimated the
contamination due to scattered light from the nuclei using identical
apertures at other nearby positions on the CCD at equal distances from the
core.  These measurements suffered more contamination than in the hotspots
so uncertainties were estimated accordingly by examining the fluctuations
in the background.  Count rates were converted to flux densities utilizing
the inverse sensitivity measurements contained in the PHOTFLAM keyword in
the image headers at the frequency given by the pivot wavelength indicated
in the PHOTPLAM keyword.  The WFPC2 filters (Holtzman et al. 1995) closely
approximate the standard optical bands: V (F606W), R (F622W, F675W,
F702W), and I (F814W). The resultant measurements are reported in
Tables~\ref{table-2}, and~\ref{table-3}, and extinction corrections were
estimated using the nearest standard band values reported in the NED
database derived from Schlegel, Finkbeiner, \& Davis (1998).

\begin{table*}
\centerline{\bf Lobe-dominated 3CR Quasar Hotspot Multi-wavelength Flux Densities}
\begin{center}
\begin{tabular}{lccc}
\hline \hline
Frequency             &Flux
&Flux                &Flux
\\
(GHz)                  &3C~208-West
&3C~275.1-North              &3C~336-South\\
\hline 
1.53                  & 0.108$\pm$0.016    & 0.355$\pm$0.053    & 0.097$\pm$0.015\\
4.89                  & 0.0277$\pm$0.003$^{a}$ & 0.178$\pm$0.027     & 0.045$\pm$0.007\\
8.44                  & 0.021$\pm$0.003    & 0.106$\pm$0.016    & 0.029$\pm$0.004\\
3.755$\times$10$^{5}$ & ...                & ...                & 0.18(0.21)$\pm$0.04$\times$10$^{-6}$\\
4.334$\times$10$^{5}$ & ...                & ...                & 0.15(0.17)$\pm$0.03$\times$10$^{-6}$\\
4.463$\times$10$^{5}$ & 0.21*(0.23)$\pm$0.04$\times$10$^{-6}$ & 0.30(0.32)$\pm$0.06$\times$10$^{-6}$ & 
... \\
4.846$\times$10$^{5}$ & ...                & ...                & 0.11(0.13)$\pm$0.03$\times$10$^{-6}$\\
4.993$\times$10$^{5}$ & ...                & ...                & 0.14(0.17)$\pm$0.05$\times$10$^{-6}$\\
\hline
\\
$\alpha_{radio}$      & 1.02$\pm$0.14*      & 0.69$\pm$0.12      & 0.74$\pm$0.12 \\
$\alpha_{ro}$         & 1.03 (1.02)*        & 1.16 (1.16)        & 1.11 (1.10) \\
\hline \hline
\end{tabular}
\caption[3CR Quasar Hotspot Fluxes]{
\label{table-2}
Fluxes are in units of Janskys with extinction corrected
values indicated in parentheses.\\
Radio spectral indices, $\alpha_{radio}$ ($F\propto\nu^{-\alpha}$) are
best-fit linear fits to the radio data.
The radio-to-optical spectra, $\alpha_{ro}$ are calculated using the 5 GHz 
and 4.463 $\times$10$^{5}$ GHz measurements for 3C~208 and 3C~275.1, 
and 4.334 $\times$10$^{5}$ GHz for 3C~336.\\
A single power law fit through all of the data points yield spectral 
indices of  1.04, 1.14, and 1.08, for the three sources, respectively.\\
$^{a}$ Taken from  \citet{bri94}.\\
The value indicated with an asterisk (*) is an optical candidate.\\
}
\end{center}
\end{table*}

\begin{table*}
\centerline{\bf 3C~454.3 Multi-wavelength Flux Densities}
\begin{center}
\begin{tabular}{lcccc}
\hline \hline
Frequency            &Flux
&Flux                &Flux
&Flux\\
(GHz)                  &Knot A
&Knot B              &Knot C 
&Hotspot\\
\hline 
1.50                  & ... & ... & ...  & 0.541$\pm$0.054 \\
4.86                  & ... & ... & ...  & 0.222$\pm$0.022 \\
8.46                  & 0.013$\pm$0.004  & 0.008$\pm$0.002 & 0.010$\pm$0.003 & 0.132$\pm$0.013 \\
22.46                 & ...    & ...   & ... & 0.065$\pm$0.010 \\
4.333$\times$10$^{5}$ & 0.25(0.32)$\pm$0.25$\times$10$^{-6}$& 0.25(0.32)$\pm$0.13$\times$10$^{-6}$& 
0.12(0.15)$\pm$0.06$\times$10$^{-6}$& 0.50(0.63)$\pm$0.10$\times$10$^{-6}$\\
\hline
\\
$\alpha_{radio}$      & ...   & ... & ...   & 0.80$\pm$0.06 \\
$\alpha_{ro}$         & 1.00 (0.98) & 0.96 (0.93) & 1.04 (1.02) & 1.15 (1.13) \\
\hline \hline
\end{tabular}
\caption[]{
\label{table-3}
Fluxes are in units of Janskys with extinction corrected 
values indicated in parentheses. \\
The radio spectral index for the hotspot, $\alpha_{radio}$ ($F\propto\nu^{-\alpha}$) is
best-fit linear fits to the radio data.
The radio-to-optical spectra, $\alpha_{ro}$ are calculated using the 8.4 GHz and 
4.333$\times$10$^{5}$ GHz measurements.
}
\end{center}
\end{table*}

\subsection{Archival Radio Data}

The quasars in which we successfully identified optical emission in the
jets and/or hotspots are well-known, and have been extensively imaged at
centimeter wavelengths, so this analysis did not warrant new observations. 
High quality radio observations were obtained from the archival databases
of the NRAO Very Large Array (Thompson et al. 1980), and additional MERLIN
observations for the three lobe-dominated quasars. The MERLIN data were
dual-frequency synthesis L-band observations which utilized a
seven-element array of Tabley, MK2, Darnhall, Wardle, Knockin, Cambridge,
and Defford. 

The list of archival radio data collected are summarized in
Table~\ref{table-1} with some effort made to find the original references.
We invariably selected 5 GHz A-configuration data (0.4\arcsec\ to
0.5\arcsec\ resolution) for each source.  The radio observations are all
to our knowledge, unpublished. In one instance (3C~454.3), we used 8.4 and
22 GHz data from one of our own recent observing programs when the source
was used as a phase calibrator for two targets. The VLA data were
calibrated using standard procedures in the NRAO AIPS reduction package
(Greisen 1988; Bridle \& Greisen 1994) and exported to the Caltech DIFMAP
program (Shepherd, Pearson, \& Taylor 1994) for imaging and deconvolution.
The initial calibration of the MERLIN data were performed at Jodrell Bank
by the archivists and then self-calibrated and imaged at Brandeis. The 1.4
and 1.7 GHz MERLIN datasets were imaged and analyzed separately and the
final images averaged. 

We used DIFMAP's MODELFIT program to model the brightness distribution in
the hotspot regions in the radio sources with elliptical gaussian
components. The inner jet in 3C~454.3 was only detected and resolved
adequately in the 8.4 GHz data so radio fluxes were measured for the jet
only at this frequency. The hotspot regions in the lobe-dominated 3CR
quasars are complex and our estimate of the errors in the extracted fluxes
are about 15$\%$. The 5 GHz flux for the 3C~208 hotspot was taken from
Bridle et al.  (1994). 

The resultant fluxes extracted in this manner are presented in
Tables~\ref{table-2} and~\ref{table-3}. The radio spectral indices
($\alpha_{\rm r}$) are linear best fits to the data and are presented
along with the radio-to-optical spectra indices ($\alpha_{\rm ro}$). In
3C~208, the radio spectrum of the hotspot is notably steeper than the
other three cases where we claim bona-fide optical detections.  The
optical flux of the optical source near the radio hotspot lies on a
power-law extrapolation of the radio spectrum (i.e., $\alpha_{\rm
r}\sim\alpha_{\rm ro}$) so it may still be due to synchrotron radiation if
the spectrum breaks at even higher energies. In the other three cases, the
radio-to-optical spectra are greater than the radio only spectra,
indicating steepening of the hotspot spectra in the observed frequency
range.

\section{Results and Descriptions of Individual Sources}

\begin{figure*}
\figurenum{1}
\begin{center}
\epsfig{file=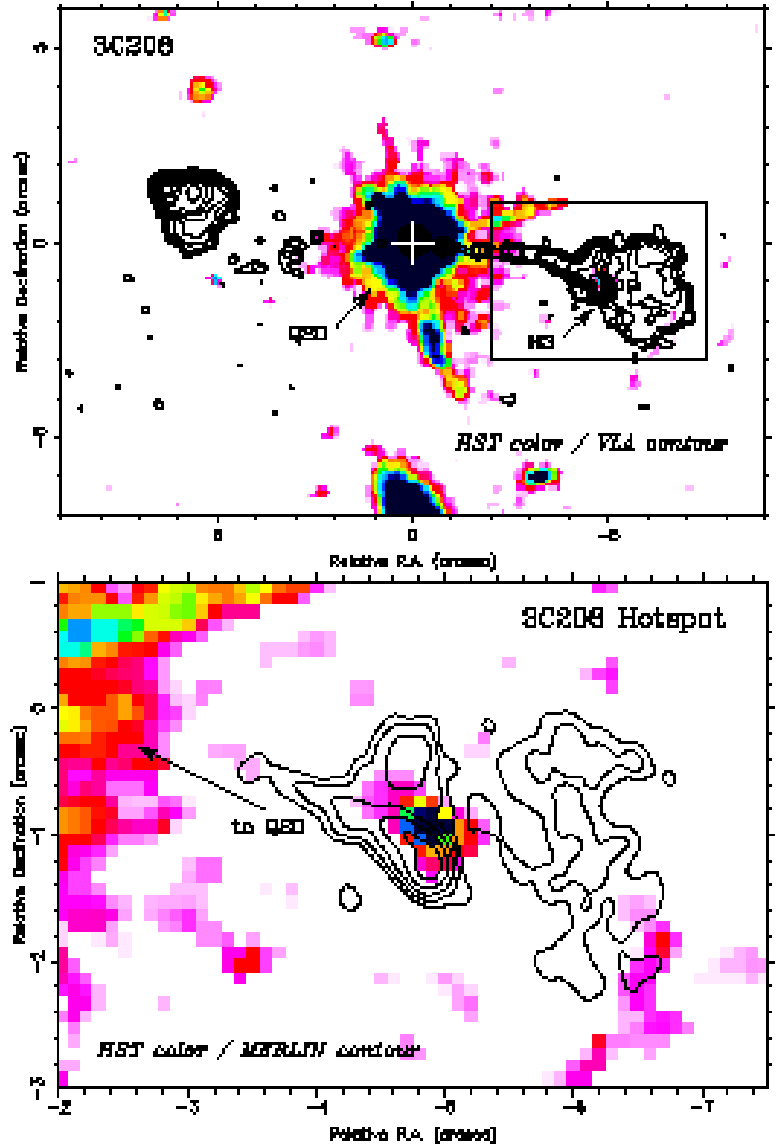,width=4.0in,angle=0}
\end{center}
\figcaption[]{\label{fig-1} Radio (contour) and optical F675W
(colorscale) overlay of 3C~208 with large field of view (top) and
zoomed in view centered near the western hotspot region (bottom panel).  The top
image shows the VLA 8.4 GHz image convolved to 0.25\arcsec\ beam. The
base level is 35$\mu$Jy/beam and spaced by factors of 2. The nucleus
is marked with a white cross. The MERLIN 1.4 GHz image on the bottom
panel has a 0.231\arcsec$\times$0.161\arcsec\ beam at PA=22$^{\circ}$
(uniform weighting). The minimum contour is 1.5 mJy/beam spaced by
factors of 2. Note that the radio/optical features in the hotspot
region are not quite aligned making this only a tentative optical
hotspot identification. North is up and east is left on the figures.}
\end{figure*}

\begin{figure*}
\figurenum{2}
\begin{center}
\epsfig{file=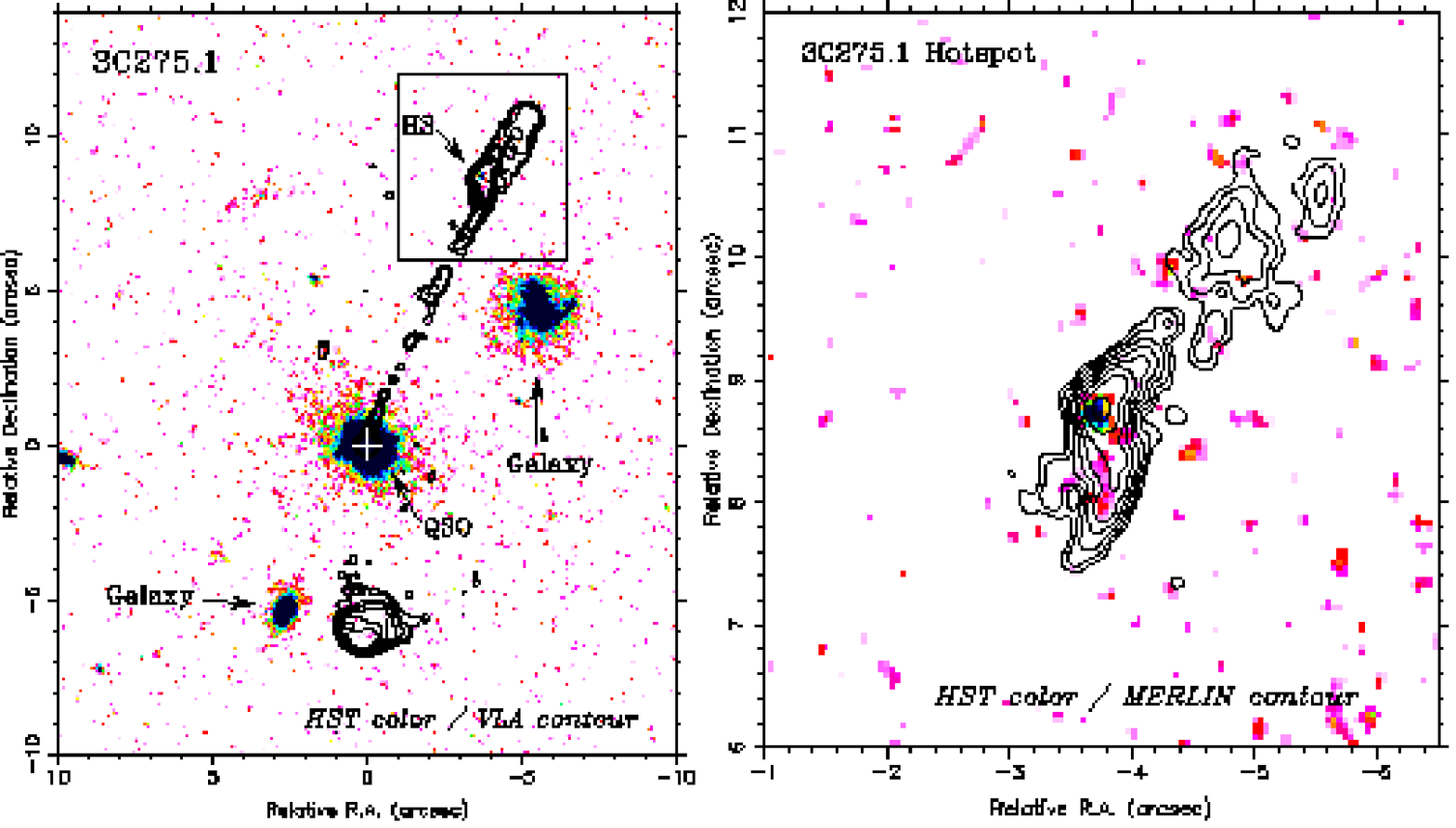,width=4.0in,angle=-90}
\end{center}
\figcaption[]{\label{fig-2} Radio (contour) and optical F675W
(colorscale) overlay of 3C~275.1 with large field of view (left) and
zoomed in view centered near the northern hotspot region (right panel). The left
image shows the VLA 8.4 GHz image convolved to 0.25\arcsec\ beam. The
base level is 0.125 mJy/beam and spaced by factors of 2 up to seven
contours to show the optical hotspot emission more clearly. The nucleus is
marked with a white cross. The MERLIN 1.4 GHz image on the right panel
has a 0.2\arcsec$\times$0.15\arcsec\ beam at PA=34$^{\circ}$ (uniform
weighting). The minimum contour is 0.75 mJy/beam spaced by factors of
2. North is up and east is left on the figures.}
\end{figure*}

\begin{figure*}
\figurenum{3}
\begin{center}
\epsfig{file=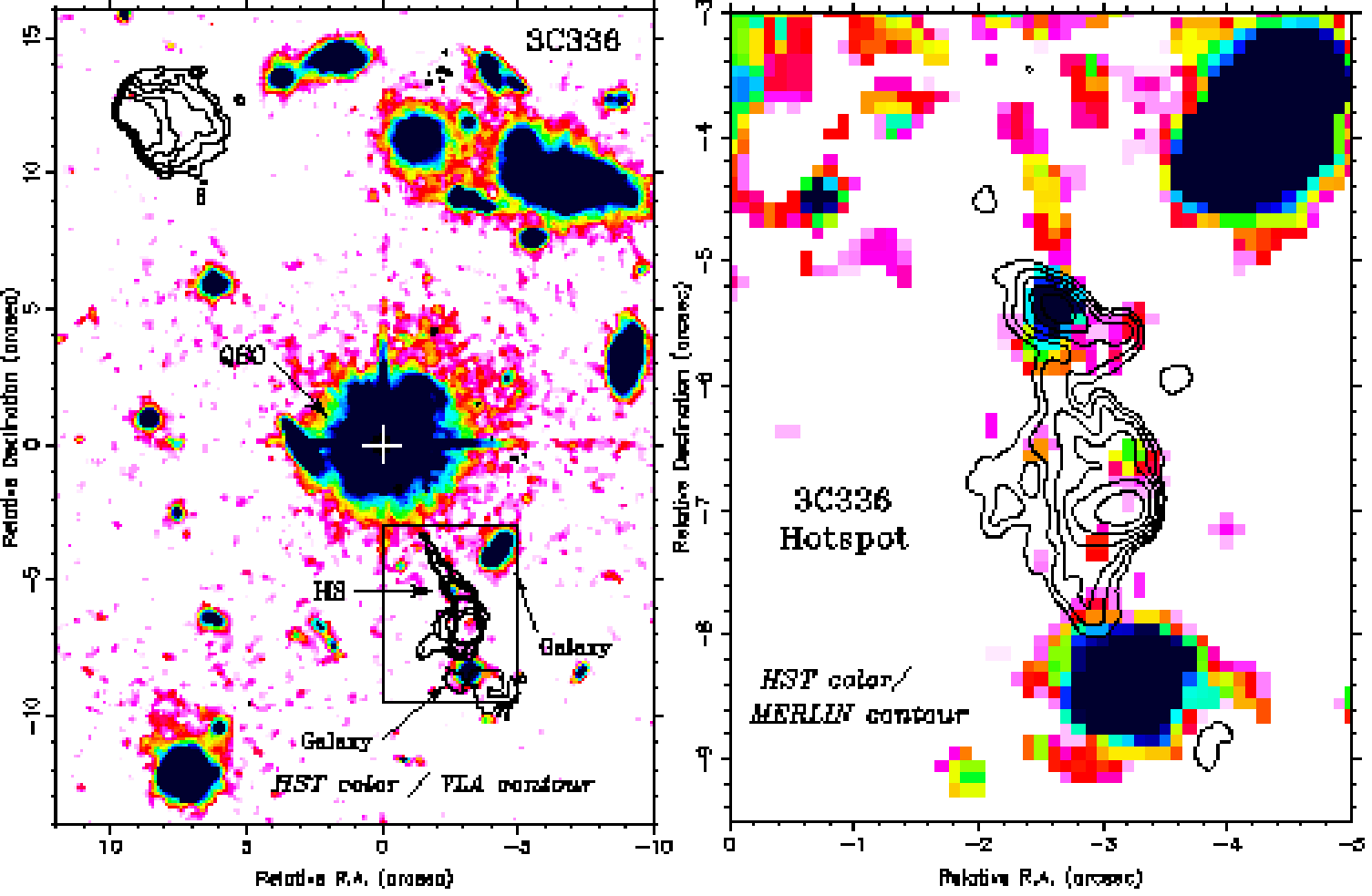,width=4.5in,angle=-90}
\end{center}
\figcaption[]{\label{fig-3} Radio (contour) and optical F702W
(colorscale) overlay of 3C~336 with large field of view (left) and
zoomed in view centered nearby the southern hotspot region (right panel).  The left
image shows the VLA 8.4 GHz image convolved with a 0.25\arcsec\
beam. The base level is 0.15 mJy/beam and spaced by factors of 2 up to
five contours to show the optical hotspot emission more clearly. The nucleus is
marked with a white cross. The MERLIN 1.6 GHz image on the right panel
has a 0.221\arcsec$\times$0.152\arcsec\ beam at PA=16.8$^{\circ}$
(uniform weighting). The minimum contour is 1 mJy/beam spaced by
factors of 2. North is up and east is left on the figures.}
\end{figure*}

We successfully identified previously unknown optical counterparts to
the hotspots of the lobe-dominated quasars 3C~275.1
(\S\ref{sec:3c275.1}; Figure~\ref{fig-2}) and 3C~336
(\S\ref{sec:3c336}; Figure~\ref{fig-3}), and in the blazar 3C~454.3
(\S\ref{sec:3c454.3}; Figure~\ref{fig-4}). We also identified a
possible optical counterpart to the hotspot in the lobe-dominated
quasar 3C~208 (\S\ref{sec:3c208}; Figure~\ref{fig-1}).  As mentioned
in the introduction, the optical hotspot detection in 3C~263
($z$=0.652) was reported since by Hardcastle et al. (2002) along
with the new Chandra X-ray detection so is not discussed here.  The
lobe-dominated quasars are best known to many as being in the Hough \&
Readhead (1989) sample of double lobed 3CR quasars. A recent summary
of their source properties and VLBI observations is in Hough et
al. (2002).

\subsection{The Double-lobed 3CR Quasars}

\subsubsection{Optical Emission Nearby the Radio Hotspot in 3C~208}
\label{sec:3c208}

The optical identification of the western radio hotspot in 3C~208
($z$=1.11) is only tentative since the optical feature is not as
precisely-aligned with any peak in the radio structure
(Figure~\ref{fig-1}) as in the other cases presented in this paper. The
closest radio feature is knot ``B'' in the nomenclature of Bridle et al.
(1994) which we identify as the hotspot. The optical feature appears
resolved; although the radio and optical peaks are
$\sim$0.3$\pm$0.1\arcsec\ apart (1.6 kpc projected), some of the optical
light may still be associated with the radio hotspot. We could not confirm
the optical detection in shallower WFPC2 F555W and F702W images
(snapshots) taken on the high resolution PC chip (Lehnert et al. 1999). 
There is no apparent optical emission from the radio jet in any of the
images. If all of the optical emission near the 3C~208 radio hotspot is
indeed associated with the radio source, its overall spectrum is
consistent with a single power-law with little or no steepening from the
observed radio spectral index of $\sim$1. Further observations with
multiple optical filters to measure the spectrum of this feature are
encouraged. 

\subsubsection{Hotspot Detection in 3C~275.1}
\label{sec:3c275.1}

3C~275.1 ($z$=0.557) is a radio source with a ``dog-leg'' appearance
(Stocke, Burns, \& Christiansen 1985). A MERLIN 1.4 GHz image appears in
Akujor et al. (1994) and a deep VLA 5 GHz image in Stocke et al. (1985). 
The optical field in 3C~275.1 has been well studied (e.g., Hintzen \&
Romanishin 1986, and references therein). 

Optical emission from the northern hotspot is prominent in the unprocessed
HST image (Figure~\ref{fig-2}) and is clearly coincident with the peak in
the high resolution MERLIN map. This detection was independently confirmed
in a recent study by Hardcastle et al. (2004).  It was never noted in
previous ground based optical imaging studies presumably due to its
faintness. There is some extended optical emission around the hotspot,
some of which is aligned along the complex radio jet/hotspot structure.
The hotspot has also been detected in the X-rays with Chandra at a flux
level above the extrapolation of the radio-to-optical spectrum signaling
an additional spectral component, presumably inverse Compton emission
(Crawford \& Fabian 2003; Hardcastle et al. 2004).

\subsubsection{Hotspot Detection in 3C~336}
\label{sec:3c336}

3C~336 ($z$=0.927) is a large angular size source, stretching across
almost 1/2 arcmin in the sky ($\sim$250 kpc projected).  It lies in an
extremely dense optical field (Steidel et al. 1997). In the nomenclature
of Bridle et al.  (1994), the optically detected feature
(Figure~\ref{fig-3}) is knot ``C'' in the jet to the south. Although it
marks the termination point of the jet, the feature does not conform to
their strict classification as a hotspot.  Rather, they classify a radio
feature further south as the hotspot although C appears more compact -- we
will proceed with our discussion assuming that C is indeed a hotspot. 

Bridle et al. (1994) found the 5 GHz flux to be 38.3 mJy compared to our
observed 45 mJy.  Our fluxes were not corrected for contribution of the
surrounding lobe emission so may be systematically higher than the fluxes
derived by Bridle et al. (1994). Region C is actually resolved into a peak
where the optical feature is clearly coincident, and a weaker extension to
the west. Our flux density measurement is from the dominant feature.
Fainter extended optical emission may be associated with the radio tail
but it is at the same level as surrounding low level emission which is
clearly not associated with the radio structure. 

We found the optical hotspot in four separate WFPC2 images obtained in
different programs (Table~\ref{table-1}). The published figure of the deep
24,000 second WFPC2 F702W exposure in Steidel et al. (1997) actually shows
the feature (at the bottom edge of their figure 2d), but was not
identified by those authors as associated with the radio jet.  This is the
only hotspot where we could measure an optical spectrum,
$\alpha_{optical}\sim$1.5$\pm$1.0, but the spectral coverage in the
optical is limited making this value highly uncertain. The difference
between the radio and optical indices is formally, $\Delta\alpha\sim$0.8.

\begin{figure*}
\figurenum{4}
\begin{center}
\epsfig{file=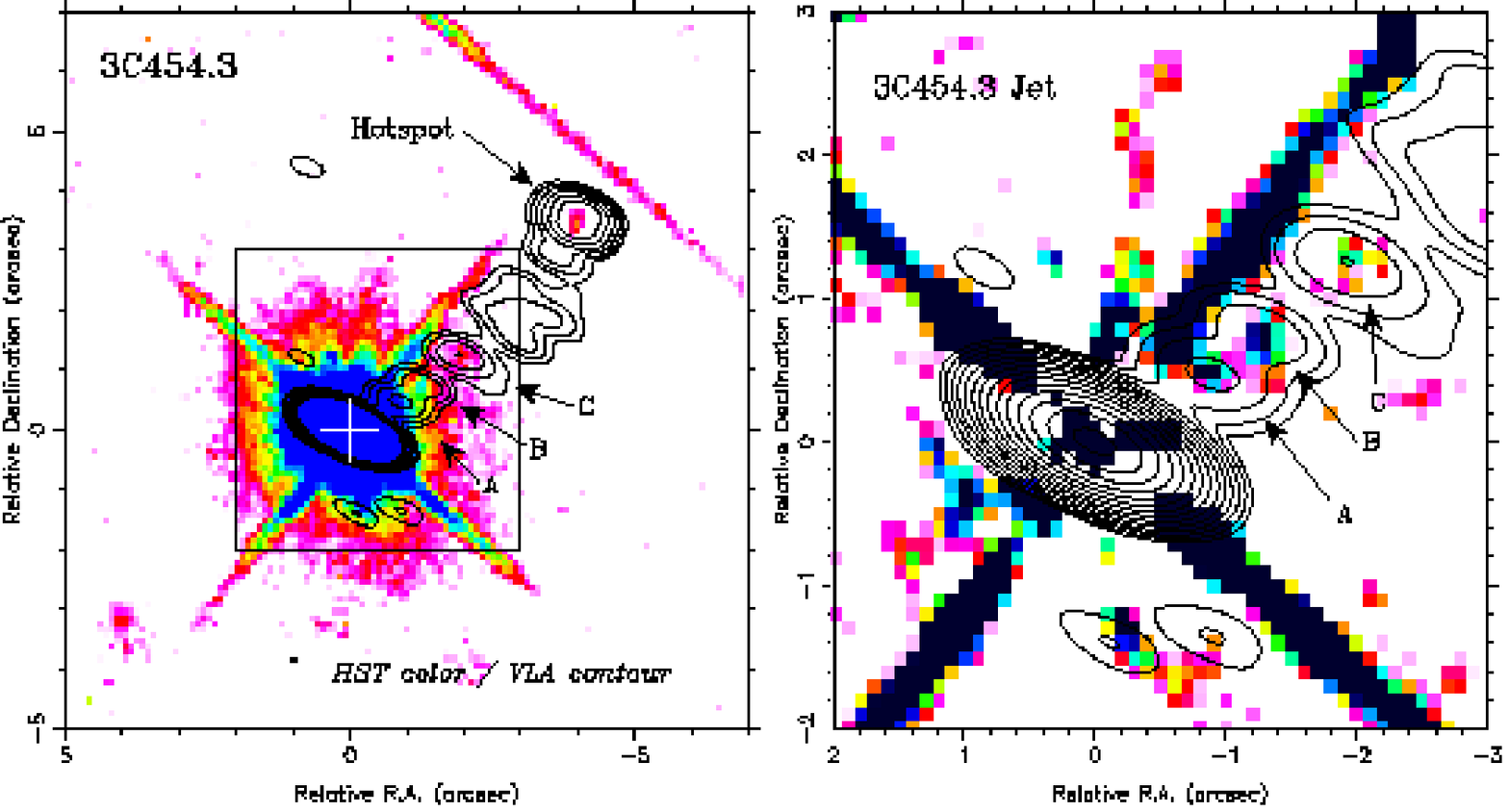,width=3.8in,angle=-90}
\end{center}
\figcaption[3c454montage.ps]{\label{fig-4} Radio (contour) and optical
HST F702W (colorscale) images of 3C~454.3 overlayed, and a zoomed in
view of the one-sided jet (right). The nucleus is marked with a white cross.
The radio image is from a VLA 8.4 GHz dataset convolved with a
beamsize of 0.7\arcsec$\times$0.3\arcsec\ at a PA=67$^{\circ}$
(uniform weighting). The base level is 1 mJy/beam and spaced by
factors of 2; only six contours are plotted on the left panel to show
the optical hotspot emission more clearly. The HST image is shown before (left)
and after (right) the subtraction of the elliptical isophote model of
the point source.  The four diffraction spikes around the optical
nucleus are clearly present, as is a portion of a spike in the upper
right of the displayed field of view from a field star (left panel).
North is up and east is left on the figures.}
\end{figure*}

\subsection{The Optical Hotspot and Jet in 3C~454.3}
\label{sec:3c454.3}

3C~454.3 ($z$=0.859) is a superluminal $\gamma$-ray blazar with a core-jet
VLBI morphology (Pauliny-Toth et al. 1987; Jorstad et al. 2001). The
parsec-scale jet stretches westward from the nucleus, continuing with more
extended emission to the northwest.  This extended emission traces a path
toward the kilo-parsec scale structure seen in VLA images, where the
optical emission was detected in the HST data (Figure~\ref{fig-4}). A
recent Chandra image detected X-ray emission from the jet and hotspot
(Marshall et al. 2005). 

A deep WFPC2 image of this quasar was obtained as part of a program to
search for Ly-$\alpha$ absorption systems in its field (Chen et al. 2001),
so the exposure was centered on one of the low resolution wide-field (WF)
CCD chips (about twice the pixel size of the higher resolution PC chip).
An unresolved (FWHM$\sim$0.15--0.2\arcsec) optical source coincident with
the position of the hotspot in the radio jet is apparent in the HST image,
very near one of the diffraction spikes from a nearby bright point source
(Figure~\ref{fig-4}; see also the unsharp-mask processed image in Figure~1
of Cheung et al.  2003). 

Upon closer inspection, a series of three bright optical knots are present
which match peaks in the jet as seen in the full resolution 8.4 GHz radio
map (a hybrid BnA configuration dataset with extra north-south
resolution). These radio features are confirmed in a MERLIN 1.7 GHz map
published in Browne et al. (1982). Our preliminary processing (Cheung et
al. 2003) of the HST data used a gaussian smoothing with a very broad
profile so that a large negative bowl suppressed emission from these inner
knots. 

The optical feature coincident with radio knot B is seen in a 280s WFPC2
PC snapshot image obtained as part of the Lehnert et al. (1999) survey of
the 3CR sample. The feature is clearly separated from the nucleus and both
its position and flux (in the same filter as in the current deep WF image)
almost match exactly. The knot is about a 3$\sigma$ detection. The fainter
outer knot (C) could not however be verified in this image as expected
from signal-to-noise limits, and knot A is simply too close to the bright
nucleus to be distinguished easily from the scattered light.  Although the
integrated flux of the hotspot is greater than knot B, it was undetected
in the higher resolution (PC chip) snapshot image. For the measured peak
surface brightness from the WF image of about 6 $\mu$Jy/arcsec$^{2}$, we
find that the expected SNR is at best about unity for the PC chip exposure
using the on-line WFPC2 exposure time calculator\footnote{See:
http://www.stsci.edu/instruments/wfpc2/}. 

The jet in the 22 GHz data was very ill-defined so fluxes for the inner
jet knots were not extracted from this dataset. However, the hotspot was
prominent.  Our total 22 GHz flux for 3C~454.3 was renormalized to
contemporaneous single dish measurements from the Mets{\"a}hovi quasar
monitoring program (H. Ter{\"a}sranta, 2003, private communication;
Ter{\"a}sranta et al. 2004). The radio source varied by at most 3$\%$ over
the 1 month period, so the average of five measurements obtained by the
Mets{\"a}hovi group was used.

\begin{table*}
\centerline{\bf Properties of the Three 3CR Quasar Hotspots}
\begin{center}
\begin{tabular}{lccc}
\hline \hline
Source  &       R (asec, 10$^{21}$ cm) & log($\nu_{\rm b}$) &  B$_{\rm eq}$   \\
(1)&(2)&(3)&(4)\\
\hline
3C~275.1-N        &0.22, 5.8&  10.0$\pm$1.2   &2.6     \\
3C~336-S          &0.25, 5.9&  11.1$\pm$1.2   &2.3     \\
3C~454.3-NW       &0.25, 6.4&  11.4$\pm$0.6   &3.8     \\
\hline \hline
\end{tabular}
\caption[]{
\label{table-4}
(2) The assumed radius of a spheroid in arcseonds (\arcsec), and in 
units of 10$^{21}$ cm.\\
(4) The logarithm of the derived break frequency, $\nu_{\rm b}$, in Hz (see \S\ref{sec:discuss}).\\
(3) Equipartition magnetic field, B$_{\rm eq}$ in units of 10$^{-4}$ G. 
These assume a uniformly filled region, negligible beaming, and 
$\gamma_{\rm min}$=10.}
\end{center}
\end{table*}

\section{Discussion}
\label{sec:discuss}

Brunetti et al.  (2003) recently found that lower power radio hotspots are
brighter at optical wavelengths than these more powerful radio hotspots
which are customarily searched for optical emission (Meisenheimer et al.
1989).  They propose a simple in-situ shock acceleration scheme whereby
high radio power hotspots with larger $B$-fields suffer greater losses,
and are inefficient at accelerating electrons up to
$\gamma$$\sim$10$^{5}$, required to produce optical emission. In this
scenario, the position of the ``break'' frequency in the observed spectrum
will depend on the balance between the rate of electron acceleration in a
shock region and combined synchrotron and inverse Compton losses. There is
a strong magnetic field dependence of the break frequency in their
synchrotron spectra, $\nu_{b}\propto$$B^{-3}$ (Figure~\ref{fig-6}), and
observations of the lower power hotspots support this prediction (Brunetti
et al. 2003;  see also earlier discussions of this scenario by
Meisenheimer et al. 1989 and Prieto et al. 2002). 

It is important to extend this work to higher power hotspots (with greater
$B$) which are expected to be faint in the near-IR and optical, and can
only be detectable in deep HST exposures. Our detections of the faint
optical counterparts in the three sources help us to constrain $\nu_{\rm
b}$ in the spectra of these high power hotspots. 

\begin{figure*}
\figurenum{5}
\begin{center}
\epsfig{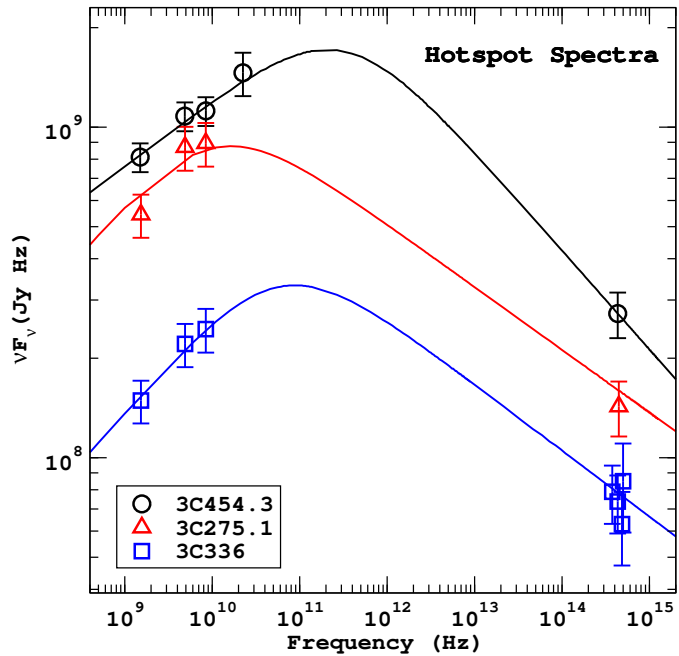}
\end{center}
\figcaption[]{\label{fig-5} Spectral energy distributions of the high
power hotspots detected in the two double-lobed quasars and the blazar
3C~454.3. Break frequencies ($\nu_{\rm b}$) are only roughly
constrained with the limited observations by fitting a double power
law spectrum to the data.}
\end{figure*}

\begin{figure*}
\figurenum{6}
\begin{center}
\epsfig{file=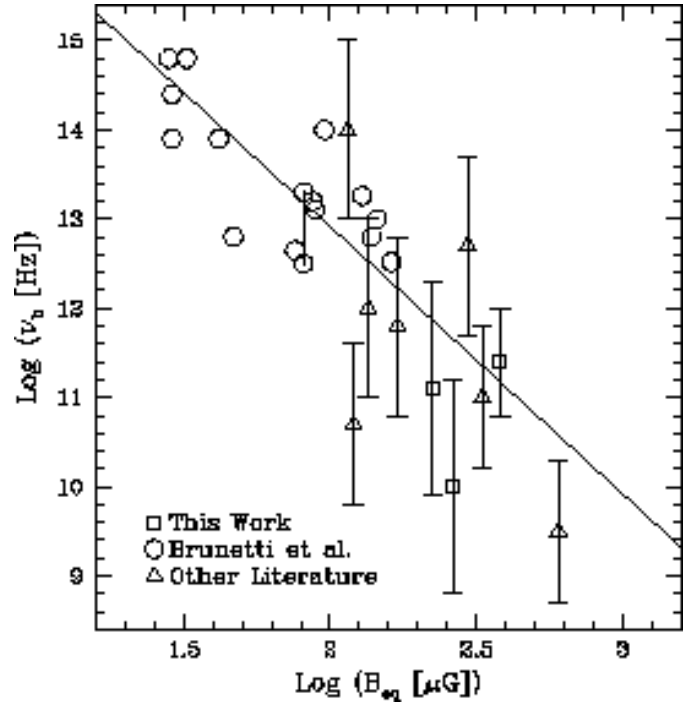,width=3.5in,angle=0}
\end{center}
\figcaption[hotspots.ps]{\label{fig-6} Break frequency ($\nu_{\rm b}$)
in the synchrotron spectra of the hotspots with optical {\it detections}
vs. magnetic field calculated assuming equipartition ($B_{\rm eq}$);
adapted from Brunetti et al. (2003).  Our new optical hotspot
detections in 3 powerful quasars (squares), and other detections from
Hardcastle et al. (2004) and Sambruna et al. (2004) (triangles) follow
approximately the $\nu_{\rm b}\propto$$B_{\rm eq}^{-3}$ trend (solid
line) expected from simple in-situ particle acceleration theory
(Brunetti et al. 2003). Uncertainties in $\nu_{\rm b}$ for the hotspots
we analyzed (Tables~\ref{table-4} \&~\ref{table-5}) are indicated with error bars. 
For the rest, the break-frequencies are within a factor of 2 as reported by Brunetti et al. 
(2003), with the exception of the case of 3C~351L where the line connecting two 
points indicates the uncertainty.}
\end{figure*}

Figure~\ref{fig-5} shows the radio-to-optical spectral energy
distributions (SED) of the three hotspots with bona-fide optical
detections from our analysis of the archival {\it HST} and radio data. The
optical fluxes clearly lie below a simple extrapolation of a power law
from the observed radio data, so are consistent with steepening in the
high energy spectrum. 

The data are admittedly sparse and the overall shape of the SED can only
be roughly constrained.  We assume that the SED can be approximated by a
simple smoothly varying double power law function so that we can
parameterize it with a break frequency ($\nu_{\rm b}$) at which the power
law slope changes.  The low energy slopes are set by the observed radio
spectra ($\alpha_{\rm r}$). In the absence of precise information on the
high energy slopes, we assume that the optical spectral index,
$\alpha_{\rm o}$=$\alpha_{\rm r}$--0.5, which is expected from first order
Fermi shock acceleration models with synchrotron losses (e.g.  Heavens \&
Meisenheimer 1987).  This is indeed, very near what is observed in the
case of the 3C~336 hotspot, where we have a rough constraint on the
optical spectrum (\S\ref{sec:3c336}).  In this manner, $\nu_{\rm b}$ is
related to observed variables by: log ($\nu_{\rm b}$) = log ($\nu_{\rm
o}$) + 2 ($\alpha_{\rm ro} - \alpha_{\rm r}$) log ($\nu_{\rm r}$/$\nu_{\rm
o}$), where the subscripts indicate the radio (5 GHz) and optical values. 
In practice, the uncertainty in determining $\nu_{\rm b}$ is dominated by
$\sigma$($\alpha_{\rm r}$), since $\sigma$($\alpha_{\rm r}$) $\sim$0.1,
compared to $\sigma$($\alpha_{\rm ro}$)$\sim$0.02--0.03. Therefore,
$\sigma$(log ($\nu_{\rm b}$)) $\sim$ 2 log ($\nu_{\rm r}$/$\nu_{\rm o}$)
$\sigma$($\alpha_{\rm r}$) $\sim$ 10 $\sigma$($\alpha_{\rm r}$).  Our
estimates of $\nu_{b}\sim10^{10}-10^{11}$ Hz for these hotspots are
therefore good to $\pm$ 1 dex (Table~\ref{table-4}). 

Following Brunetti et al. (2003), we used the observed radio spectra, flux
densities at 5 GHz, and constraints on the sizes from our modelfits, to
calculate magnetic fields assuming equipartition (Table~\ref{table-4}).
These magnetic fields are in the range of a few hundred $\mu$G, so are
relatively large.  As in Brunetti et al. (2003), we have assumed for
simplicity that the hotspot emission is not relativistically beamed (but
see Georganopoulos \& Kazanas 2003). If the radiation is beamed, the
equipartition magnetic fields calculated must be reduced by a factor of
$\delta$$^{5/7}$ to transform to the source rest-frame (Stawarz, Sikora,
\& Ostrowski 2003), where $\delta$ is the Doppler factor.

\begin{table*}
\centerline{\bf Optical Hotspots from the Literature}
\begin{center}
\begin{tabular}{lccccccc}
\hline \hline
Source    &z      &$\alpha_{\rm r}$ &Ref.  & $\alpha_{\rm ro}$ & Ref.  & log($\nu_{\rm b}$)  & B$_{\rm eq}$\\
(1)&(2)&(3)&(4)&(5)&(6)&(7)&(8)\\
\hline
0405--123-N &0.574  &0.90$\pm$0.10 & C04; S04; T05  &1.10  &C04; S04; T05  &12.7$\pm$1.0 & 2.9\\
3C~228-S    &0.5524 &0.76*         & J95       &1.05  &H04       &11.8$\pm$1.0 & 1.7\\
3C~245-W    &1.029  &0.87$\pm$0.08 & C04; S04; T05  &1.39  &C04; S04; T05  & 9.5$\pm$0.8 & 6.0\\
1136--135-W &0.554  &0.85$\pm$0.08 & C04; S04; T05  &1.22  &C04; S04; T05  &11.0$\pm$0.8 & 3.3\\
1150+497-H  &0.334  &0.72$\pm$0.09 & C04; S04; T05  &1.12  &C04; S04; T05  &10.7$\pm$0.9 & 1.2\\
1150+497-I  &0.334  &0.80$\pm$0.10 & C04; S04; T05  &1.07  &C04; S04; T05  &12.0$\pm$1.0 & 1.4\\
3C~403-F1   &0.059  &0.8*          & D99       &0.87  &H04       &14.0$\pm$1.0 & 1.1\\
3C~403-F6   &0.059  &0.8*          & D99       &0.85  &H04       &14.1$\pm$1.0 & 1.4\\
\hline \hline
\end{tabular}
\caption[]{
\label{table-5}
$^{*}$Errors of 0.1 assumed.\\
(3) Radio spectral index from references in column (4).\\
(5) Radio-to-optical spectral index from references in column (6).
We used fluxes reported by the authors at 5 GHz in the radio (in 3C~228 and 3C~403,
H04 converted from 8.4 GHz using $\alpha_{\rm r}$=0.5), and
3.45$\times$10$^{14}$ (3C~228), 4.34$\times$10$^{14}$ (3C~403), and
5.13$\times$10$^{14}$ Hz (rest) for the optical.\\
(7) The logarithmic of the derived break frequency, $\nu_{\rm b}$, in Hz (see \S\ref{sec:discuss}).\\
(8) Equipartition magnetic field, B$_{\rm eq}$ in units of 10$^{-4}$ G. See Table~\ref{table-4} 
for assumptions used.\\
References in columns (4) and (6):
C04 = Cheung (2004), Cheung et al. in prep.;
D99 = Dennett-Thorpe et al. (1999);
H04 = Hardcastle et al. (2004);
J95 = Johnson et al. (1995);
L92 = Liu et al. (1992);
S04 = Sambruna et al. (2004);  
T05 = Tavecchio et al. (2005)
}
\end{center}
\end{table*}

Recent work by Sambruna et al. (2004) and Hardcastle et al. (2004) provide
additional optical hotspot source detections and we were able to estimate
($\nu_{\rm b}$, $B_{\rm eq}$) for 8 additional hotspots in 6 radio sources
as described above (Table~\ref{table-5}). Although information for a large
number of optical hotspots are provided by Hardcastle et al. (2004), only
four of these hotspots (in three objects) were not previously considered
in the study by Brunetti et al. (2003). Of these four hotspots, we could
not confirm the case in 3C~280 from our independent analysis of the same
HST data (see also the radio/optical image overlay in Figure~3 of Ridgway
et al. 2004 -- the radio and optical peaks were not clearly coincident as
in our case in 3C~208; \S~\ref{sec:3c208}) -- this case is omitted in our
discussion. 

We find that the derived ($\nu_{b}$, $B$) of the high radio power hotspots
(Tables~\ref{table-4} \&~\ref{table-5}) lie near the extrapolation of the
$\nu_{\rm b}\propto$$B^{-3}$ trend traced by Brunetti et al. (2003), which
lends support to their simple picture (Figure~\ref{fig-6}).  This is a
remarkable finding considering the hotspots span about 3 orders of
magnitude in magnetic field energy density, while considering both radio
galaxies and quasars.  In the scenario outlined by Brunetti et al. (2003),
the proportionality is: $\nu_{\rm b}\propto$ $(u_{l}/u_{\rm HS})^{1/2}
\tau^{-2} B^{-3}$, where $\tau$ is the dynamical age of the hotspot and
the ratio of the energy densities of the lobe ($u_{l}$) and the hotspot
($u_{\rm HS}$) accounts for adiabatic losses. More precise mapping of the
high-power hotspot SEDs are necessary to confirm that the --3 slope indeed
extends over 6 orders of magnitude in frequency (e.g. that there is no
change in the slope). If confirmed, these observations could imply that
that there is a single dynamical age for radio hotspots in radio galaxies
and quasars over a wide range in power. 

Spitzer Space Telescope observations, already approved for two of our
new optical hotspot sources, will allow us to measure the high energy
slope of the SED which will constrain $\nu_{b}$ more robustly.  While
additional millimeter and sub-mm observations (in the future with
ALMA) are required to map the shape of the high energy synchrotron
spectrum, existing radio, optical, and new infrared data can constrain
the break frequency adequately (to better than a decade) for the
purpose of showing the extrapolation of the $\nu_{\rm b}-B_{\rm eq}$
sequence determined from the theory.

\acknowledgements

\noindent
{\bf Acknowledgements}

Perl scripts written by Dan Homan were essential for this work and we are
grateful to him for providing them. We thank Harri Ter{\"a}sranta for
providing us Mets{\"a}hovi monitoring data in advance of publication,
Samuel Hariton for his help in the early stages of this project, the
staffs at NRAO and Jodrell Bank for supplying us the archival radio data,
and the anonymous referee for useful comments.  C.~C.~C. is grateful to
the HETG group at the MIT Kavli Institute for hosting his fellowship. 

Radio astronomy at Brandeis University is supported by the NSF through
grant AST 00-98608. Further support to C.~C.~C. and J.~F.~C.~W. came
from NASA grant GO2-3195C from the Smithsonian Astrophysical
Observatory, and HST-GO-09122.08-A from the Space Telescope Science
Institute (STScI). 

The VLA is a facility of the 
National Radio Astronomy Observatory is operated by Associated
Universities, Inc. under a cooperative agreement with the National
Science Foundation (NSF).
MERLIN is a National Facility operated by the University of Manchester
at Jodrell Bank Observatory on behalf of PPARC.
Based on observations made with the NASA/ESA Hubble Space Telescope,
obtained from the data archive at the STScI.  STScI is operated by the
Association of Universities for Research in Astronomy, Inc. under NASA
contract NAS 5-26555.
This research has made use of NASA's Astrophysics Data System Abstract
Service and the NASA/IPAC Extragalactic Database which is operated by
the Jet Propulsion Laboratory, California Institute of Technology,
under contract with the NASA.

\begin{thebibliography}{}

\bibitem[Akujor et al.(1994)]{aku94} Akujor, C.~E., L{\" u}dke, E.,
Browne, I.~W.~A., Leahy, J.~P., Garrington, S.~T., Jackson, N., \&
Thomasson, P.\ 1994, \aaps, 105, 247

\bibitem[Blandford \& Eichler(1987)]{bla87}
Blandford, R.~\& Eichler, D.\ 1987, Phys. Rep., 154, 1

\bibitem[Bridle \& Greisen(1994)]{bri94a} Bridle, A.~H. \& Greisen, E.~W.  
1994, AIPS Memo 87 (NRAO: Charlottesville)

\bibitem[Bridle et al.(1994)]{bri94} Bridle, A.~H., Hough, D.~H.,
Lonsdale, C.~J., Burns, J.~O., \& Laing, R.~A.\ 1994, \aj, 108, 766

\bibitem[Browne et al.(1982)]{bro82} Browne, I.~W.~A., Clark, R.~R.,
Moore, P.~K., Muxlow, T.~W.~B., Wilkinson, P.~N., Cohen, M.~H., \&
Porcas, R.~W.\ 1982, \nat, 299, 788


\bibitem[Brunetti et al.(2003)]{bru03} Brunetti, G., Mack, K.-H.,
Prieto, M.~A., \& Varano, S.\ 2003, \mnras, 345, L40

\bibitem[Catanese \& Weekes(1999)]{cat99}  
Catanese, M.~\& Weekes, T.C. 1999, \pasp, 111, 1193

\bibitem[Celotti, Ghisellini, \& Chiaberge(2001)]{cel01} Celotti, A.,
Ghisellini, G., \& Chiaberge, M.\ 2001, MNRAS, 321, L1

\bibitem[Chartas et al.(2000)]{cha00} Chartas, G., Worrall, D.~M.,
Birkinshaw, M.,~et al.\ 2000, \apj, 542, 655

\bibitem[Chen et al.(2001)]{che01} Chen, H.-W., Lanzetta, K.~M., Webb,
J.~K., \& Barcons, X.\ 2001, \apj, 559, 654

\bibitem[Cheung(2002)]{che02} Cheung, C.~C.\ 2002, \apjl, 581, L15

\bibitem[Cheung et al.(2003)]{che03} Cheung, C.~C., Wardle, J.~F.~C.,
Chen, T., \& Hariton, S.~P.\ 2003, New Astronomy Review, 47, 423

\bibitem[Cheung(2004)]{che04} Cheung, C.~C.\ 2004, Ph.D. Thesis, Brandeis 
University

\bibitem[Crawford \& Fabian(2003)]{cra03} Crawford, C.~S.~\& 
Fabian, A.~C.\ 2003, \mnras, 339, 1163 

\bibitem[Dennett-Thorpe et al.(1999)]{den99} Dennett-Thorpe, J.,
Bridle, A.~H., Laing, R.~A., \& Scheuer, P.~A.~G.\ 1999, \mnras, 304,
271

\bibitem[Georganopoulos \& Kazanas(2003)]{geo03} Georganopoulos, M., \& 
Kazanas, D.\ 2003, \apj, 589, L5

\bibitem[Greisen(1988)]{gre88} Greisen, E.~W. 1988, AIPS Memo 61 (NRAO: 
Charlottesville)

\bibitem[Hardcastle et al.(2002)]{hard02} Hardcastle, M.~J., Birkinshaw,
M., Cameron, R.~A., Harris, D.~E., Looney, L.~W., \& Worrall, D.~M.\ 2002,
\apj, 581, 948

\bibitem[Hardcastle et al.(2004)]{hard04} Hardcastle, M.~J., Harris, D.~E., 
Worrall, D.~M., \& Birkinshaw, M.\ 2004, \apj, 612, 729

\bibitem[Heavens \& Meisenheimer(1987)]{hea87} Heavens, A.~F. \&
Meisenheimer, K. 1987, \mnras, 225, 335

\bibitem[Hintzen \& Romanishin(1986)]{hin86} Hintzen, P.~\& Romanishin,
W.\ 1986, \apjl, 311, L1

\bibitem[Holtzman et al.(1995)]{hol95} Holtzman, J.~A., Burrows, C.~J.,
Casertano, S., Hester, J.~J., Trauger, J.~T., Watson, A.~M., \& Worthey,
G.\ 1995, \pasp, 107, 1065

\bibitem[Hough \& Readhead(1989)]{hou89} Hough, D.~H.~\& Readhead,
A.~C.~S.\ 1989, \aj, 98, 1208

\bibitem[Hough et al.(2002)]{hou02} Hough, D.~H., Vermeulen, 
R.~C., Readhead, A.~C.~S., Cross, L.~L., Barth, E.~L., Yu, L.~H., Beyer, 
P.~J., \& Phifer, E.~M.\ 2002, \aj, 123, 1258 

\bibitem[Johnson, Leahy, \& Garrington(1995)]{joh95} Johnson, R.~A.,
Leahy, J.~P., \& Garrington, S.~T.\ 1995, \mnras, 273, 877

\bibitem[Jorstad et al.(2001)]{jor01} Jorstad, S.~G., Marscher, A.~P.,
Mattox, J.~R., Wehrle, A.~E., Bloom, S.~D., \& Yurchenko, A.~V.\ 2001,
\apjs, 134, 181

\bibitem[Le Brun et al.(1997)]{leb97} Le Brun, V., Bergeron, J., Boisse,
P., \& Deharveng, J.~M.\ 1997, \aap, 321, 733

\bibitem[Lehnert et al.(1999)]{leh99} Lehnert, M.~D., Miley, G.~K.,
Sparks, W.~B., Baum, S.~A., Biretta, J., Golombek, D., de Koff, S.,
Macchetto, F.~D., \& McCarthy, P.~J.\ 1999, \apjs, 123, 351

\bibitem[Liu \& Zhang(2002)]{liu02} Liu, F.~K.~\& Zhang, Y.~H.\ 2002,
\aap, 381, 757

\bibitem[Liu, Pooley, \& Riley(1992)]{liu92} Liu, R., Pooley, G., \&
Riley, J.~M.\ 1992, \mnras, 257, 545

\bibitem[Marshall et al.(2003)]{mar03} Marshall, H.~L., Cheung, T., 
Canizares, C.~R., \& Fang, T.\ 2003, BAAS, 36, 4.26

\bibitem[Marshall et al.(2005)]{mar05} Marshall, H.~L., Schwartz,
D.~A., Lovell, J.~E.~J., et al. 2005, ApJS, 156, 13

\bibitem[Meisenheimer et al.(1989)]{mei89} Meisenheimer, K., R{\"o}ser,
H.-J., Hiltner, P.~R., Yates, M.~G., Longair, M.~S., Chini, R., \& Perley,
R.~A.\ 1989, \aap, 219, 63

\bibitem[Meisenheimer, Yates, \& R{\"o}ser(1997)]{mei97} Meisenheimer, K.,
Yates, M.~G., \& R{\"o}ser, H.-J.\ 1997, \aap, 325, 57

\bibitem[Pauliny-Toth et al.(1987)]{pau87} Pauliny-Toth, I.~I.~K., Porcas,
R.~W., Zensus, J.~A., Kellermann, K.~I., \& Wu, S.~Y.\ 1987, \nat, 328,
778

\bibitem[Prieto, Brunetti, \& Mack(2002)]{pri02} Prieto, 
M.~A., Brunetti, G., \& Mack, K.\ 2002, Science, 298, 193 

\bibitem[Ridgway \& Stockton(1997)]{rid97} Ridgway, S.~E.~\& Stockton, A.\
1997, AJ, 114, 511

\bibitem[Ridgway, Stockton, \& Lacy(2004)]{rid04} Ridgway, S.~E.,
Stockton, A., \& Lacy, M. 2004, \apj, 600, 70

\bibitem[Sambruna et al.(2004)]{sam04} Sambruna, R.~M., 
Gambill, J.~K., Maraschi, L., Tavecchio, F., Cerutti, R., Cheung, C.~C., 
Urry, C.~M., \& Chartas, G.\ 2004, \apj, 608, 698 

\bibitem[Schlegel, Finkbeiner, \& Davis(1998)]{sch98} Schlegel, D.~J.,
Finkbeiner, D.~P., \& Davis, M.\ 1998, \apj, 500, 525

\bibitem[Schwartz et al.(2000)]{sch00} Schwartz, D.~A., Marshall, H.~L.,
Lovell, J.~E.~J.~et al.\ 2000, \apjl, 540, L69

\bibitem[Shepherd, Pearson, \& Taylor(1994)]{she94} Shepherd, M.~C.,
Pearson, T.~J., \& Taylor, G.~B. 1994, BAAS, 26, 987

\bibitem[Stawarz, Sikora, \& Ostrowski(2003)]{sta03} Stawarz, \L., 
Sikora, M., \& Ostrowski, M. 2003, \apj, 597, 186

\bibitem[Steidel et al.(1997)]{ste97} Steidel, C.~C., Dickinson, M., Meyer,
D.~M., Adelberger, K.~L., \& Sembach, K.~R.\ 1997, \apj, 480, 568

\bibitem[Stocke, Burns, \& Christiansen(1985)]{sto85} Stocke, J.~T.,
Burns, J.~O., \& Christiansen, W.~A.\ 1985, \apj, 299, 799

\bibitem[Tavecchio et al.(2000)]{tav00} Tavecchio, F., Maraschi, L.,
Sambruna, R.~M., \& Urry, C.~M.\ 2000, \apjl, 544, L23

\bibitem[Tavecchio et al.(2005)]{tav05} Tavecchio, F., Cerutti, R., Maraschi, L.,
Sambruna, R.~M., Cheung, C.~C., \& Urry, C.~M.\ 2005, \apj, submitted

\bibitem[Ter{\" a}sranta et al.(2004)]{ter04} Ter{\"  a}sranta, H., et
al.\ 2004, \aap, 427, 769

\bibitem[Thompson et al.(1980)]{tho80} Thompson, A.~R., Clark, B.~G.,
Wade, C.~M., \& Napier, P.~J.\ 1980, \apjs, 44, 151

\bibitem[Veron-Cetty \& Veron(1996)]{ver96} Veron-Cetty M.-P. \& Veron
P. 1996, ESO Scientific Report 17

\end{thebibliography}
\end{document}